\documentclass{elsart}

\usepackage{epsfig}

\usepackage{graphicx}

\newcommand{\marge}[1]{\marginpar{}}  
\newcommand{\Sl}[1]{{}}           
\newcommand{\beq}[1]{\Sl{#1}\begin{equation}\if#1\empty\else\label{#1}\fi}
\newcommand{\eeq}{\end{equation}}
\newcommand{\beqa}[1]{\Sl{#1}\begin{eqnarray}\if#1\empty\else\label{#1}\fi}
\newcommand{\eeqa}{\end{eqnarray}}

\newcommand{\Eq}[1]{Eq.(\ref{#1})}

\begin{document}
\begin{frontmatter}

\title{Nonextensive statistics in viscous fingering}

\author{Patrick Grosfils, Jean Pierre Boon\thanksref{CAICYT}}
 
\address{Center for Nonlinear Phenomena and Complex Systems\\ 
Universit\'e Libre de Bruxelles, 1050 - Bruxelles, Belgium}
\thanks[CAICYT]{{\em E-mail address} : {\tt jpboon@ulb.ac.be}; 
{\em URL} : {\tt http://poseidon.ulb.ac.be}}

\begin{abstract}
Measurements in turbulent flows have revealed that the 
velocity field in nonequilibrium systems exhibits 
$q$-exponential or power law distributions in agreement with 
theoretical arguments based on nonextensive statistical mechanics. 
Here we consider Hele-Shaw flow as simulated by the Lattice 
Boltzmann method and find similar behavior from the analysis 
of velocity field measurements. For the transverse velocity, 
we obtain a spatial $q$-Gaussian profile and a power 
law velocity distribution over all measured decades. To explain
these results, we suggest theoretical arguments based on 
Darcy's law combined with the non-linear advection-diffusion
equation for the concentration field. Power law and $q$-exponential
distributions are the signature of nonequilibrium systems with 
long-range interactions and/or long-time correlations, and therefore 
provide insight to the  mechanism of the onset of fingering processes.  

\bigskip



\end{abstract}
\end{frontmatter}

Experiments in turbulent flows (such as Couette-Taylor flow 
\cite{swinney}) have shown that data obtained from
quantities extracted from velocity field measurements
in these nonequilibrium systems exhibit power law or
$q$-exponential velocity distributions which have been analyzed 
and interpreted with theoretical arguments based on nonextensive 
statistical mechanics \cite{beck}. Other physical systems have been 
shown to exhibit similar type distributions as a consequence of
{\it nonextensivity}  \cite{tsallis}. Here we consider fingering 
which occurs in the interfacial zone between two fluids confined 
between two plates with a narrow gap (Hele-Shaw geometry) when  
one of the fluids is displaced by the other fluid, the interfacial instability
resulting from mobility differences between the fluids.
Using a mesoscopic approach - the lattice Boltzmann method - we 
investigated the dynamics of spatially extended Hele-Shaw flow
\cite{grosfils}. We are primarily interested in the early stage
of the fingering process in order to explore the characteristics
of the velocity field after the destabilization of the interface.

The Hele-Shaw geometry is intrinsically tri-dimensional, 
but the effective destabilization of the flow can be described in the 
2-D plane, the flow being purely Poiseuille flow in the third
dimension of the shallow layer \cite{darcy}. So we can simulate Hele-Shaw 
flow by using the two-dimensional LBGK equation, the Bhatnagar-Gross-Krook 
version of the lattice Boltzmann equation where the full collision term
is approximated by a single relaxation term \cite{succi}. One emulates
3-D flow by introducing a drag term, thereby simulating a system with 
a virtual cell gap in the third dimension: the drag enters the 
LBGK equation as a damping term \cite{flekkoy}.

As the simulation method has been described in detail in a previous 
publication  \cite{boek}, we merely outline the essentials of the physics
of the problem as it is simulated. The system consists of a 2-D
rectangular box with vertical length $L_y$ and horizontal width 
$L_x$ ($L_x \times L_y = 1024 \times 2048$ nodes) filled with fluid 2 
which is being displaced by fluid 1 injected uniformly from the top of 
the box (the respective drag coefficients are $\beta_1 < \beta_2$). 
Fluid 1 invades the system through a constant drift applied along the 
$y$-direction. The initial concentration profile of fluid 1 is a step function
which, because of mutual diffusion between the two miscible fluids,
develops into an erfc $y$ profile. The interfacial profile which is 
initially flat along the transverse $x$-direction, is perturbed with
white noise to trigger the instability. Any small perturbation so
induced is the siege of a local pressure gradient normal to the
interface where from the instability develops as illustrated in Fig.1.

Measurements of the time evolution of the mixing length ($L_{mix}$) 
of the interfacial zone (see Fig.1) show a dynamical transition from the 
short time diffusive regime to the nonlinear regime \cite{boek}. 
The short time behavior obeys a power law $L_{mix}\propto t^{\mu}$ 
with $\mu \simeq 1/2$; then (as shown in Fig.3 in \cite{boek}) a transition is 
observed from the diffusive regime with the ``early time exponent'' $1/2$ 
to the regime with a ``dynamic exponent'' $\mu \simeq 2.3$, 
a value which was obtained from lattice Boltzmann simulation as well 
as from direct numerical simulations of Darcy's equation \cite{boek}. 
Here we shall restrict to the investigation of the transverse velocity
field ($v_x$), that is the field transverse to the propagation direction of 
the flow, during the onset of fingering. What is meant  by {\em onset} 
is the late stage of the diffusive regime with $\mu \simeq 1/2$ (see Fig.3
in  \cite{boek}) where the fingers have developed into a regular pattern 
(see Fig.1) before growing into the competitive stage where large fingers 
absorb slower smaller fingers, a feature characteristic of the nonlinear 
regime with $\mu \simeq 2.3$.

When the planar interface destabilizes, the flow develops a wiggling 
concentration profile (in the $x$-direction) thereby producing local 
concentration gradients which trigger vorticity fluctuations. 
Pattern formation follows in the velocity field as shown in Fig.2a, and 
``vortices'' are being created with alternating polarities. They are 
reminiscent of vortex patterns in two-dimensional turbulent flows, although 
here the Reynolds number has low value. However the relevant control 
parameter, the P\'eclet number, is high: $Pe = L_x |{\bf v}_y|/D \simeq 200$ 
($D$ is the diffusion coefficient). We performed measurements of the
velocity field and found that the transverse component $v_x$ in each 
vortex shows a $q$-Gaussian profile (see Fig.2b), and that the corresponding 
velocity distribution, $P(v_x)$, obtained by computing $v_x$ over all vortices,  
follows a power law, $P(v_x) \propto v_x^{-0.875}$ (see Fig.3). 

It has been shown that processes stemming from the generalized
Fokker-Planck equation \cite{tsallis,bukman,borland} or the
generalized advection-diffusion equation \cite{malacarne,pedron}
can produce stationary solutions of the $q$-exponential type provided 
some conditions are met for the drift and dissipative functions. 
In the present simulations, we observe that the concentration field
exhibits a $q$-Gaussian profile which suggests that the concentration
$C(x,t)$ is governed by the "porous media equation" \cite{compte}

\beq{a1}
\partial_t C(x,t) + \partial_x [K(x)\,C(x,t)]\,=\,\partial_x [D(x)\, \partial_x C^{\alpha} (x,t)]\,,
\label{PME}
\eeq 
where $K(x)$ and $D(x)$ are the drift and dissipation functions respectively. 
Note that the second term in (\ref{a1}) can be rewritten formally as
\beq{a2}
\partial_x [D\, \partial_x C^{\alpha}]\,=\,\partial_x [D_{\alpha} \,\partial_x C]\,,
\label{D_q}
\eeq 
where $D_{\alpha}\,=\,\alpha \,D_0\,C^{\alpha -1}$.
The stationary solution to Eq.(\ref{a1}) is a $q$-exponential  of the form
(see e.g. \cite{borland})
\beq{a3}
C(x) = C_0\, e_q^{-\gamma_q \, |x|^{\lambda}} 
= C_0 \,[1-(1-q)\gamma_q \, |x|^{\lambda}]^{\frac{1}{1-q}}\,,
\label{q-G-C_x}
\eeq 
where $C_0$ is a normalization constant, and  $q+\alpha = 2$. The concentration 
profile found from the simulation results has exactly this form with $\lambda = 2$.

On the other hand, the velocity field is related to the concentration through
Darcy's law (see e.g. \cite{homsy}) which can be written as
\beq{a4}
\nabla \times [\beta (C)  {\bf v}]\,=\,0\,,
\label{Darcy}
\eeq 
or 
\beq{a5}
R\,({\bf v} \times \nabla C)\,+\,(\nabla \times {\bf v})\,=\,0\,,
\label{Darcy2}
\eeq 
with $R\,=\,\partial \ln \beta / \partial C$, where
$\beta$ is the damping function which, in the fingering simulations, 
controls the drag and depends on the concentration $C(x)$.
The solution to Eq.(\ref{a5}) is a function ${\bf v}(C)$
which, through the concentration, is space dependent, but is
analytically unknown. However, given that the concentration exhibits
a $q$-Gaussian profile (\ref{a3}), one can expect that the velocity field
also assumes a $q$-Gaussian form. For the transverse velocity field, 
Fig.2b shows that the $q$-Gaussian interpretation is in perfect
agreement with the data obtained from the simulation results, i.e.
\beq{a6}
v_x (x) =  v_0\, [1-(1-q)\phi_q \, x^2]^{\frac{1}{1-q}}\,,
\label{q-G-v_x}
\eeq 
where $v_0$ is a normalization constant. Note that the value of
$q$ in (\ref{a6}) is not the same as in (\ref{a3}).
So the pattern of alternating vortices of the transverse velocity 
amplitude as seen in Fig.2a can be viewed as a landscape of hills and 
wells of radial $q$-Gaussians distributed along the concentration 
profile of the interfacial zone.

We next consider the velocity distribution which follows from this
spatial profile. The $q$-Gaussian spatial profile (shown in Fig.2b)
is obtained experimentally by measuring the $v_x$ amplitude in a 
section plane orthogonal to the $y$-axis. Now each $q$-Gaussian
"blob" in Fig.2a is a function of $r^2 = x^2 + y^2$ and is expected
 to be radially symmetric. So generalizing the $q$-exponential in 
 \Eq{q-G-v_x} to an arbitrary power $\lambda$ of the radial variable 
 $r$, we have
 \beq{a7}
v(r) \equiv v_x(r)/ v_0\,=\,e_q^{-\phi_q \, r^\lambda}\,=\, 
[1-(1-q) \phi_q r^\lambda]^{\frac{1}{1-q}}\,,
\label{q-G-v_r}
\eeq 
We seek to establish the corresponding 
probability distribution function $P(v)$ which, for the sake of
generality, is considered in $d$ dimensions
\beq{a8}
P(v)\,=\, c_d \, \int_0^\infty r^{d-1} |dr|  \,\delta(v(r)-v)\,,
\label{pdf_v}
\eeq
where $c_d$ is an integration constant.
Expression (\ref{q-G-v_r}) is inverted to give 
$r^{\lambda}= - \frac{1}{\phi_q}\,\ln_q v$ 
which is used in (\ref{pdf_v}) to obtain
\beq{a9}
P(v) \,=\, c_d \,  \frac{|r|^{d-\lambda} v^{-q}} {\lambda \,\phi_q}
       \,=\, \frac{c_d}{\lambda \,\phi_q^{d/\lambda}}
       \left[- \ln_q \,v \right] ^\frac{d-\lambda}{\lambda} v^{-q}\,.
\label{pdf_v_q}
\eeq 
For two-dimensional systems ($d=2$) with $q$-Gaussian spatial profile
($\lambda = 2$), we obtain the power law distribution
\beq{a10}
P(v)\,= \,\frac{c_d}{\phi_q} \, v^{-q}\,.
\label{pdf_power}
\eeq 
This is exactly the power law that we find for the transverse velocity distribution 
as exhibited in Fig.3  where a fit to the simulation data gives an exponent $q < 1$
(indicating statistical importance of large values of the velocity).

In conclusion, we have presented  an analysis and an interpretation 
of the statistical behavior of the transverse velocity field in the "onset" region 
of the fingering process just below the dynamical transition.
To the best of our knowledge, this transition has remained unexplained 
analytically. The velocity distribution analysis developed in
the present paper offers a method to investigate both the 
transverse and longitudinal velocity fields, $P(v_x)$ and $P(v_y)$
respectively, below and above the dynamical transition, and 
should therefore provide insight in the nature of the transition as will
be discussed elsewhere.

\ack{This work was supported by a grant from the 
{\em European Space Agency} and {\em PRODEX} (Belgium) 
under contract ESA/14556/00/NL/SFe(IC).}

\begin{figure}

\begin{center}
\rotatebox{+90}{
\resizebox{8cm}{!}
{
\includegraphics*[5cm,0cm][18cm,25cm]{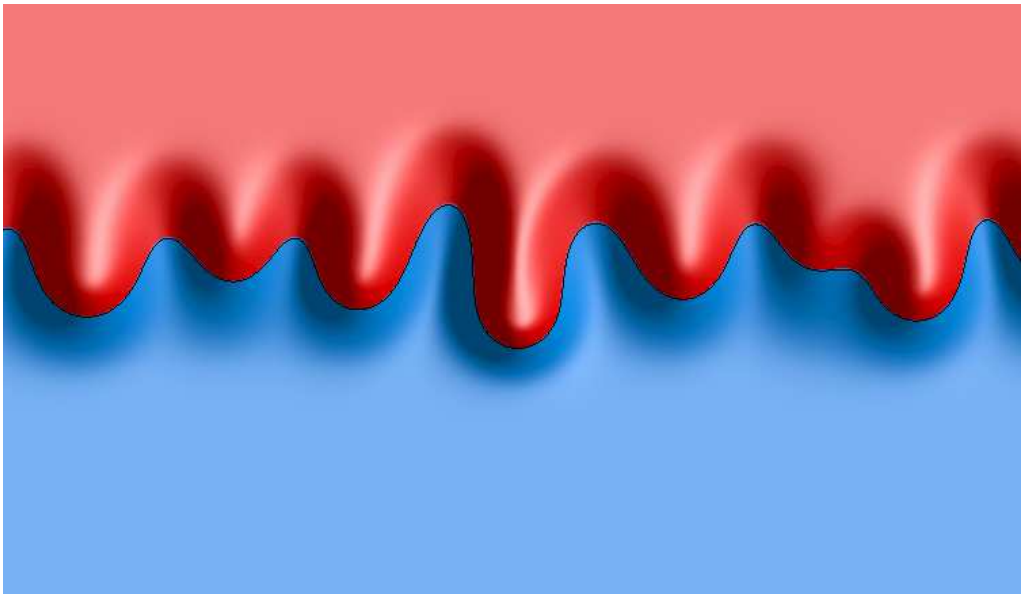}
}
}

\caption{Illustration of Lattice Boltzmann (BGK) simulation of viscous fingering. 
               The mixing length between the two fluids is clearly identified by the 
                shadow zone around the density ($=0.5$) isoline
               (color code indicating the concentration 
                of the invading fluid from red ($C_1=1$) to blue ($C_1=0$)).}
\end{center}  
\label{fig_fingering}

\end{figure}

\begin{figure}

\begin{center}
\rotatebox{-90}{
\resizebox{9cm}{!}{
\includegraphics*[0cm,0cm][18cm,25cm]{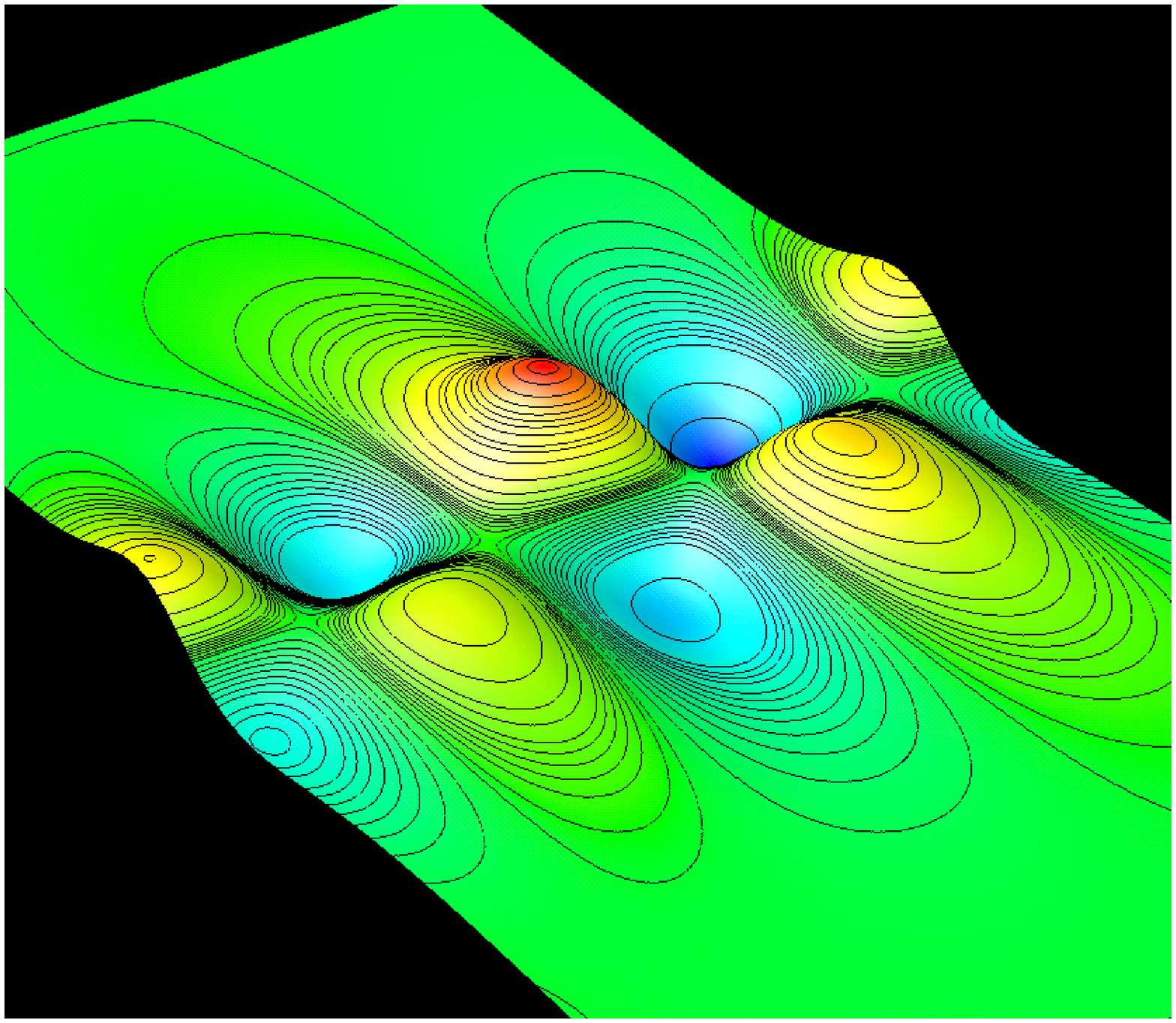}
}
}

\rotatebox{-90}{
\resizebox{9cm}{!}{
\includegraphics*[2cm,0cm][21cm,25cm]{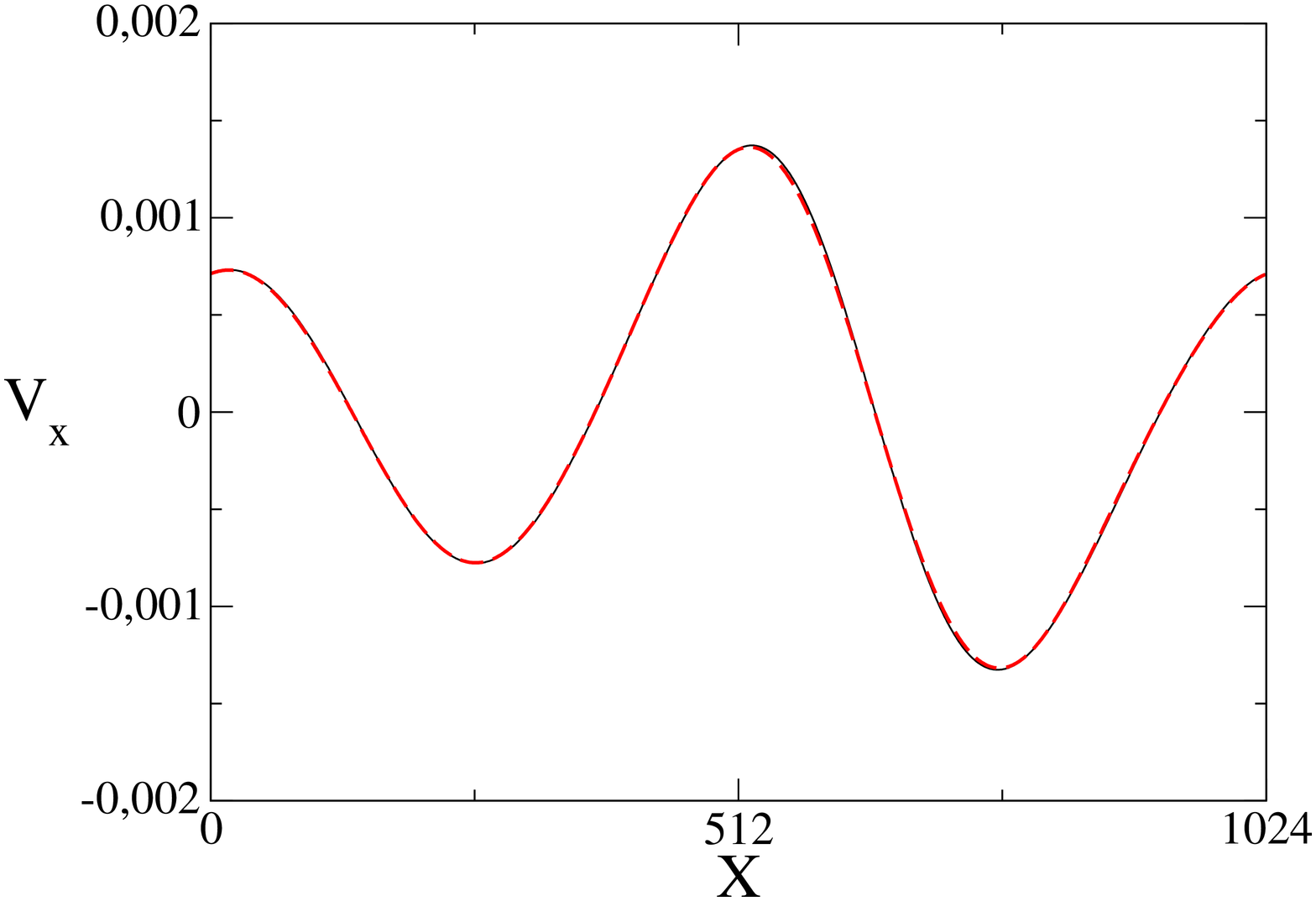}
}
}

\caption{(a) Upper panel:
Transverse velocity  field showing landscape of $q$-Gaussian 
"hills and wells"; simulation data with iso-velocity  contours,
color code indicating highest positive values (red) to largest 
negative values (blue) of $v_x$.
(b) Lower panel: Transverse velocity profile (solid line) 
obtained from the simulation data by a section plane  cut through 
the hills and wells parallel to and North-West of the main valley in 
the upper panel; analytical function Eq.(\ref{a6}) (dashed line): 
the convex and concave curves are connected by summation of  
$q$-Gaussians. }
\end{center}  
\label{fig_v_profile}

\end{figure} 

\begin{figure}
\begin{center}
\resizebox{12cm}{!}{
\rotatebox{-90}{
\includegraphics{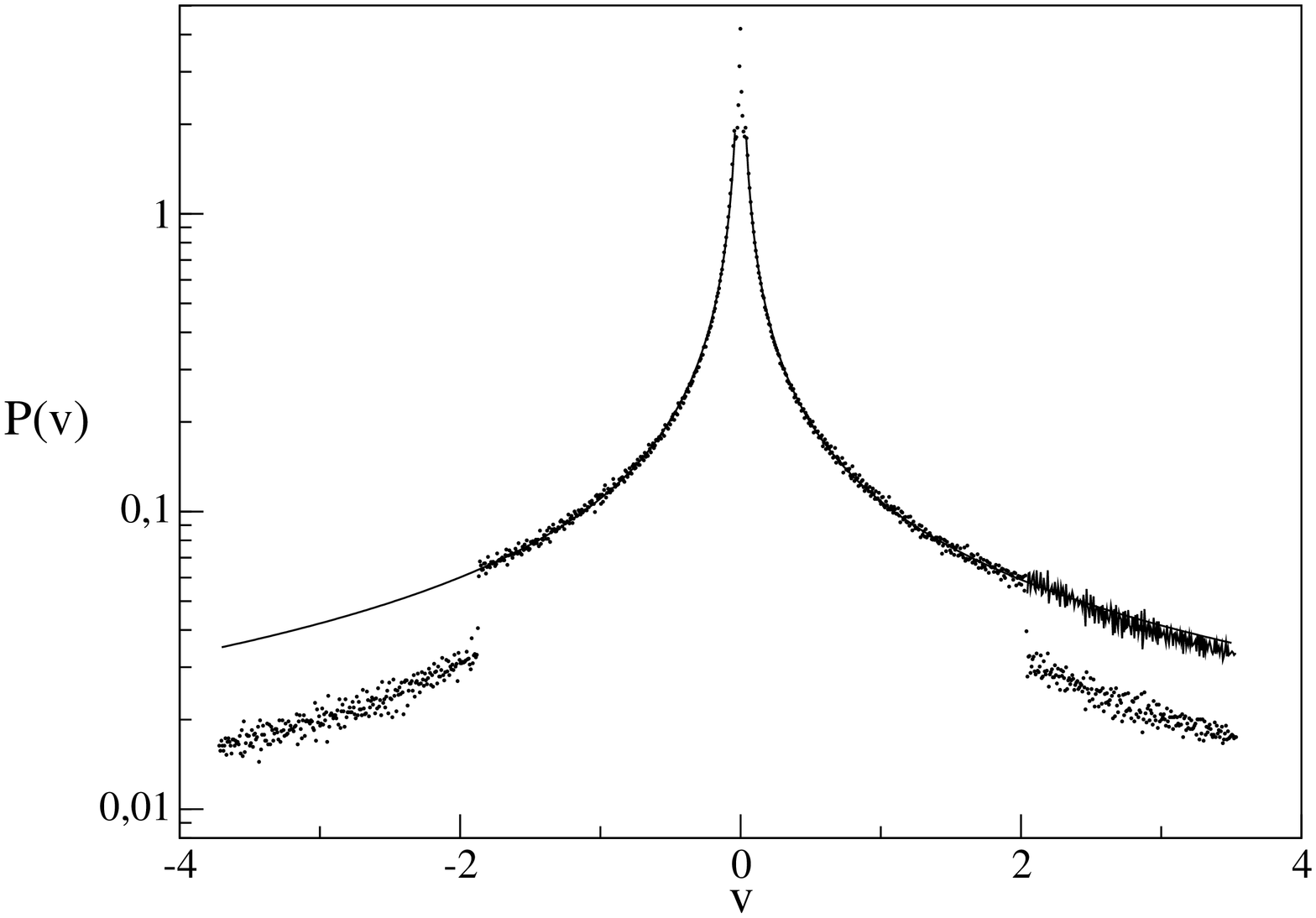}}}
\resizebox{12cm}{!}{
\rotatebox{-90}{ 
\includegraphics{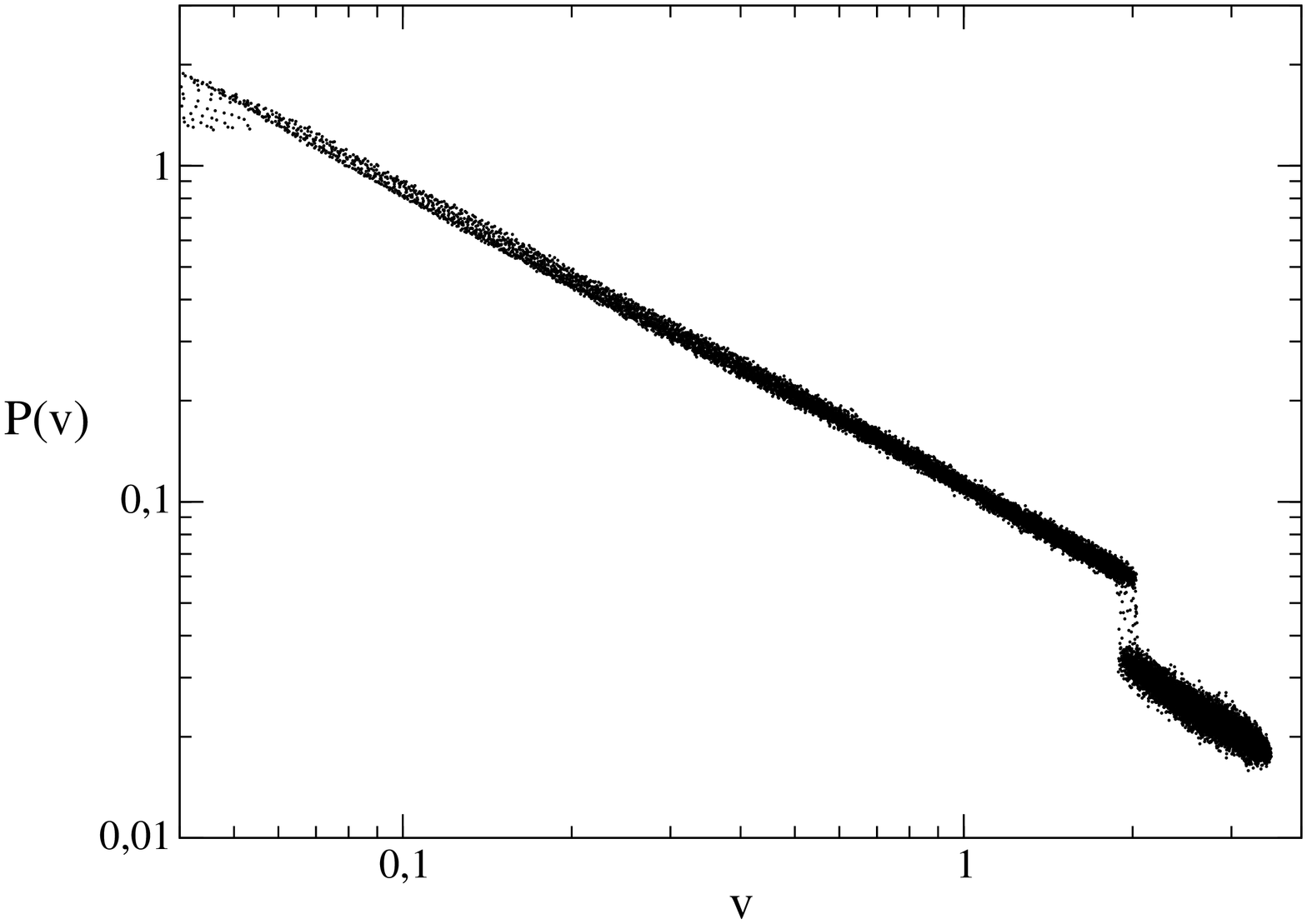}}}
\caption{Velocity distribution with power law fit : $P(v)$ with 
$v=(v_x - \langle v_x \rangle) \,\sigma^{-1}$ where
$\langle v_x \rangle$ is the mean transverse velocity and
$\sigma = [\langle v_x^2 \rangle - \langle v_x \rangle ^2]^{1/2}$.
Data are rescaled (shifted upwards in the far right branch of the upper panel) 
showing that all data follow the same power law down to values 
$v \simeq 0.05$ (below which statistics become insufficient; 
here 5 points with $v < 0.05$ out of 1000).
$P(v) = A |v| ^{-q}$ (Eq.(\ref{a10})) with $q = 0.8739$ and $A = 0.1082$ 
for $|v > 0|$, and  $q = 0.8787$ and $A = 0.1054$ for $|v < 0|$. Lower 
panel illustrates power law for 40 sets of data taken at successive times.}
\end{center}
\label{fig_v_distribution}
\end{figure}


\begin{thebibliography}{99}

\bibitem{swinney}
C. Beck, G.S. Lewis, and H. L. Swinney,
{\it Phys. Rev.E}  (2001) {\bf 63}, 035303(R).

\bibitem{beck} 
C. Beck,
{\it Phys. Rev. Lett.} (2001) {\bf 87}, 180601.
 
\bibitem{tsallis}
C. Tsallis,
{\it Physica D} (2004) {\bf 193}, 3. 

\bibitem{grosfils}
P. Grosfils and J.P. Boon,
{\it J. Mod. Phys. B} (2003) {\bf 17}, 15 .

\bibitem{darcy}
See e.g. J. Bear, {\it Dynamics of Fluids in Porous Media}
(Dover, New York, 1988).

\bibitem{succi}
S. Succi, {\it The Lattice Boltzmann Equation for Fluid 
Dynamics and Beyond} (Clarendon Press, Oxford, 2001).

\bibitem{flekkoy}
E.G. Flekkoy, U. Oxaal, J. Feder, and T. Jossang,
{\it Phys. Rev.E}  (1995) {\bf 52}, 4952.

\bibitem{boek}
P. Grosfils, J.P. Boon, J. Chin, and E.S. Boek, 
{\it Phil. Trans. Royal Soc.} (2004) {\bf 362}, 1723.

\bibitem{bukman}
C. Tsallis and D.J. Bukman,
{\it Phys. Rev.E} (1996) {\bf 54}, R2197.

\bibitem{borland}
L. Borland,
{\it Phys. Lett. A} (1998) {\bf 245}, 67; 
{\it Phys. Rev.E}   (1998) {\bf 57}, 6634.

\bibitem{malacarne}
L.C. Malacarne, R.S. Mendes, I.T. Pedron, and E.K. Lenzi,
{\it Phys. Rev.E}  (2001) {\bf 63},  030101R.

\bibitem{pedron} 
I.T. Pedron, R.S. Mendes, L.C. Malacarne, and E.K. Lenzi,
{\it Phys. Rev.E}  (2002) {\bf 65},  0411108.

\bibitem{compte}
A. Compte and D. Jou,
{\it J. Phys. A: Math. Gen.} (1996) {\bf 29}, 4321.

\bibitem{homsy}
G.M. Homsy, 
{\it Ann. Rev. Fluid Mech.} (1987) {\bf 19}, 271.

\end{thebibliography}
\end{document}